\documentstyle[12pt]{article}
\advance\textheight by \topskip
\begin{document}
\thispagestyle{empty}
\begin{flushright}
SU--ITP--95--8\\
IEM--FT--86/95\\
SUSSEX--AST--95/4--2\\
gr-qc/9504022\\
\today
\end{flushright}
\vskip 1cm
\begin{center}
{\LARGE\bf Stationary solutions in Brans--Dicke
\vskip 2mm
stochastic inflationary cosmology}
\vskip 1cm

{\bf Juan Garc\'{\i}a--Bellido}\footnote{
E-mail: j.bellido@sussex.ac.uk}
\vskip .1mm
Astronomy Centre, School of Mathematical and Physical Sciences, \\
University of Sussex, Brighton BN1 9QH, UK.
\vskip 3mm
{\bf Andrei Linde}\footnote{
E-mail: linde@physics.stanford.edu}
\vskip .1mm
Department of Physics, Stanford University, \\
Stanford, CA 94305--4060, USA
\end{center}
\vskip .5cm

{\centerline{\large\bf Abstract}}
\begin{quotation}
\vskip -0.4cm
In Brans--Dicke theory the Universe becomes divided after inflation
into many exponentially large domains with different values of the
effective gravitational constant. Such a process can be described by
diffusion equations for the probability of finding a certain value of
the inflaton and dilaton fields in a physical volume of the Universe.
For a typical chaotic inflation potential, the solutions for the
probability distribution never become
stationary but grow forever towards larger values of the fields. We show
here that a non-minimal conformal coupling of the inflaton to the curvature
scalar, as well as radiative corrections to the effective potential, may
provide a dynamical cutoff and generate stationary solutions.
We also analyze the possibility of large nonperturbative jumps of the
fluctuating inflaton scalar field, which was recently revealed in the context
of the Einstein theory. We find that in the Brans--Dicke theory the amplitude
of such jumps is strongly suppressed.

\end{quotation}
\newpage

\baselineskip=17.5pt
\section{\label{1}Introduction}

After fifteen years of development of inflationary cosmology,
the basic principles of this theory seem to be well understood,
see e.g. Ref.~\cite{Book}.  However, there is still a very long way
from these basic principles to the final theory. One of the main
problems is the absence of the final version of the underlying
particle theory. Fortunately, many properties of inflationary
cosmology are very stable with respect to the change of its
particular realization. In particular, most inflationary models
predict a flat Universe ($\Omega = 1$) with a nearly scale-independent
spectrum of density perturbations. Still some important deviations from
the standard lore may appear when one goes from one theory of elementary
particles to another. For example, in a certain class of theories, one
can obtain an open \cite{Open} or a closed \cite{Lab,OMEGA} Universe, or
even a Universe consisting of different  causally disconnected regions
with all possible values of $\Omega < 1$ \cite{OMEGA}.
It would be important to find out whether some
other modifications of the inflationary paradigm may appear when one
implements it in different theories of fundamental interactions.

An interesting playground for testing the robustness of various ideas
about inflation is provided by the  Brans--Dicke theory \cite{JBD}.
It remains to be seen whether this theory or some of its
generalizations can have a sufficiently good motivation,
e.g. from the point of view of string theory
\cite{Olive}--\cite{Polyakov}. In any case, however,  some
qualitatively new effects which appear in the  BD inflationary
cosmology   may justify its investigation.

One of these effects is the possibility to avoid the graceful
exit problem of the first-order inflation~\cite{EI}. This
possibility, however, requires the introduction of an effective
potential and/or a nonminimal kinetic term for the Brans--Dicke
field~\cite{HEI}, which makes the corresponding theory rather
complicated and may lead to certain problems~\cite{LiddleLyth}.
In some of the recent versions of these models
the end of inflation occurs in the standard way, during the
stage of slow rolling~\cite{CrittStein}.

In what follows we will be interested in another specific effect
which may appear in the inflationary Brans--Dicke cosmology. In
this theory quantum fluctuations of the Brans--Dicke field
$\phi$ during inflation driven by the inflaton field $\sigma$
lead to the division of the Universe into different
exponentially large regions where the effective gravitational
constant $G = M_{\rm P}^{-2} ={\omega\over 2\pi \phi^2}$ can
take all possible values from $0$ to $\infty$, see Ref.~\cite{ExtChaot}.

This effect becomes especially interesting if one takes into
account the process of self-repro\-duction of inflationary
universe. This process can be studied in a most adequate way by
using the stochastic approach to inflation and investigating
the probability distribution $P_p(\phi,\sigma,t)$, which gives the
relative fraction of the volume of the Universe containing the
inflaton field $\sigma$ and the Brans--Dicke field $\phi$
at the moment $t$. A detailed study of
the distribution $P_p$ was performed recently in Ref.~\cite{LLM},
where it was shown that in many inflationary models based on
the Einstein theory of gravity (i.e. the theory with a  constant
Brans--Dicke field) this probability distribution
rapidly approaches a stationary regime. This means that if one
takes a section of the Universe at a given time $t$ and
calculates the relative fraction of domains of the Universe with
given properties, the result will not depend on the time $t$,
both during inflation and after it.

However, the stationary character of the probability
distribution $P_p$ in general relativity  is closely related to
the existence of the Planck boundary, where the potential energy
density $V(\sigma)$ becomes comparable with the Planck density
$M_{\rm P}^4$.  Typically the distribution $P_p(\sigma,t)$
rapidly moves towards large $\sigma$, for the reason that the
volume of domains with large $V(\sigma)$ grows very fast. The
distribution $P_p(\sigma,t)$ becomes stabilized as it approaches
the Planck boundary, where, as it is argued in
\cite{LLM}--\cite{GBL}, the process of self-reproduction of
inflationary domains is less efficient, or inflation becomes
impossible altogether.

Meanwhile, in the Brans--Dicke theory the situation is more
complicated~\cite{GBLL,JGB}. Since the effective Planck mass  in
this theory depends on $\phi$,\, $M_{\rm P}^2(\phi) =
{2\pi\over\omega}\,\phi^2$,
the Planck boundary instead of being a point becomes a
line $\phi^4 = {\omega^2\over4\pi^2}\,V(\sigma)$. Then the
distribution $P_p(\phi,\sigma,t)$ first moves towards the Planck
boundary, and after that it slides along this boundary
in the direction where the rate of expansion of the Universe
becomes greater.  Such runaway solutions typically lead to the
distributions $P_p$ rapidly moving towards indefinitely large
values of $\phi$ and $\sigma$. As a consequence, a typical value
of the Planck mass at the end of inflation also becomes
indefinitely large,   and the effective gravitational constant
vanishes.

In fact, runaway solutions may appear even in the ordinary
Einstein theory.  This may happen if the inflaton field is
nonminimally coupled to curvature due to the existence of the
interaction   $\ -{1\over 2} \xi\sigma^2 R$, with
 $\xi < 0$~\cite{GBL}. On the other hand, in the theories with
$\xi >0$ one may encounter many different regimes. Depending on
the value of $\xi$, one may not have inflation at all, or one
may have inflation without self-reproduction of inflationary
domains, or inflation and  self-reproduction with a stationary
distribution $P_p$.

This suggests us to   study   a more general version of the
Brans--Dicke theory, where not only the Brans--dicke field $\phi$,
but also the inflaton field $\sigma$ is nonminimally coupled to
gravity. Some models of this type have been discussed before in
\cite{Hybrid,Liddle}. Note that this framework seems to be much
more natural than the standard one, where it is assumed that
only one of the two scalar fields (the Brans--Dicke field) is
nonminimally coupled to gravity.

This paper is organized as follows.  In Section~\ref{2} we
discuss the problem of runaway solutions in the Brans--Dicke
stochastic inflation. In Section~\ref{3} we derive a set of
equations for inflation in the Brans--Dicke theory with a non
minimally coupled inflaton scalar field. We study the case of a
purely massive scalar field and find that it provides a natural
cutoff for the motion of the inflaton and dilaton fields,
resulting in the existence of stationary solutions to the
diffusion of both fields. In order to test the
stability of the results, we analyse the same model in the
presence of a non zero self-coupling of the inflaton. We find
that such a term is enough to destroy the stationarity of the
probability distribution, making such a stationarity rather
improbable. In Section~\ref{4} we propose a different solution
to the runaway behavior of the scalar fields by considering
1--loop corrections to the effective potential. For certain
values of the parameters of the model, it is possible that
the effective potential of the inflaton field acquires a maximum
at exponentially large values of $\sigma$. This  provides a natural
cutoff for the rate of inflation and makes the distribution $P_p$
stationary. Furthermore, the Planck mass in this model is
exponentially larger than the only scale in the problem, the
mass of the inflaton, thus naturally explaining a hierarchy of
scales. In Section~\ref{5} we explore the
possibility of very large quantum jumps of the scalar field
\cite{OPEN} in the context of the Brans-Dicke theory.
In Section~\ref{6} we draw some conclusions.

\section{\label{2}Runaway solutions in Brans--Dicke inflation}

In this section we will introduce the problem of runaway solutions
in Brans--Dicke cosmology. For a detailed analysis see Refs.
\cite{GBLL,JGB}. Consider the evolution of the inflaton field
$\sigma$ with a generic chaotic potential in a JBD theory of
gravity with dilaton field $\phi$,
\begin{equation}\label{SP}
{\cal S}=\int d^4x \sqrt{-g} \left[{1\over8\omega} \phi^2 R -
{1\over2}(\partial\phi)^2 - {1\over2}(\partial\sigma)^2 -
V(\sigma)\right]\ ,
\end{equation}
where Planck mass is written in terms of the dilaton field as
$M_{\rm P}^2(\phi) = {2\pi\over\omega} \phi^2$. For generic inflaton
potentials of the type $V(\sigma) = {\lambda\over 2n}\sigma^{2n}$, the
equations of motion for the homogeneous fields in the slow-roll
approximation are
\begin{equation}\label{SRE}
{\dot\phi\over\phi} = {H\over\omega}, \hspace{1cm}
{\dot\sigma\over\sigma} = - {n\over2}\,{H\over\omega}\,
{\phi^2\over\sigma^2}\ ,
\end{equation}
This implies that
\begin{equation}\label{QQ}
{d\over dt} \left(\phi^2 + {2\over n}\sigma^2\right) = 0 \ ,
\end{equation}
which means that the fields move along a circle of radius $r^2 =
\phi^2 + \varphi^2$, with $\varphi^2 ={2\over n}\,\sigma^2$, see
Ref.~\cite{ExtChaot}.  Inflation begins at the Planck
boundary   $\phi^2 = {\omega\sqrt\lambda\over 2\pi\sqrt 2n}\,
\sigma^n$ and ends somewhere at the line    $\phi^2 \simeq {\omega\over
2}\,\sigma^2 \gg \sigma^2$.  This implies that if inflation begins at
$\sigma_0 \gg \phi_0$, it will end at $\phi \sim \sqrt{2\over n} \sigma_0$.
Thus, if we find a reason why $\sigma_0$ should be large, we will be
able to explain why $M_{\rm P} = \sqrt{2\pi\over\omega}\phi \sim 2
\sqrt{\pi\over n\omega}\sigma_0$ is also large.

This simple picture becomes much more complicated if one takes into
account that in addition to the classical motion, the fields $\phi$ and
$\sigma$ experience quantum jumps with a typical amplitude ${H\over
2\pi} $ during each time interval $H^{-1}$. One can easily verify
that (for $\phi \ll \sigma$ )
these jumps are greater than the classical rolling of the fields for
\begin{equation}\label{PP}
{3\pi^2\over \omega^3} \phi^4 < V(\sigma)
< {4 \pi^2 \over \omega^2} \phi^4 \ ,
\end{equation}
the last inequality corresponding to the  Planck boundary. Thus, if the field
is not very far away from the Planck boundary (but still sufficiently far away
so that our approach remains reliable), the motion of the field occurs not due
to classical rolling but due to Brownian jumps in all possible directions.  The
jumps in the direction of greater Hubble constant lead to much more rapid
expansion of space. As a result, very soon (in the synchronous time $t$) the
main part of the physical volume of the Universe shifts towards the Planck
boundary where the Hubble constant takes its greatest values. After that the
field $\sigma$ can take all possible values along the Planck boundary, and
therefore the value of the field $\phi$ in the domains reaching the boundary of
the end of inflation does not depend on the initial value $\sigma_0$.

 However, the Hubble constant $H(\phi,\sigma)$ is different in different parts
of the Planck boundary. If it has a maximum   at  some point  $\sigma =
\sigma_{\rm max}$ along the Planck boundary, then the main part of the volume
of
the Universe will be produced as a result of exponentially rapid inflation near
this point. Therefore the main part of the volume of the Universe after
inflation will be produced with $M_{\rm P} \sim \sqrt{4\pi\over
n\omega}\,\sigma_{\rm max}$~\cite{GBLL}.

Unfortunately, as we have shown in~\cite{GBLL,JGB}, this regime is not realized
in the   model (\ref{SP}) with $V(\sigma) = {\lambda\over 2n}\sigma^{2n}$, nor
in the models with exponentially growing potentials. The reason is very
simple: The Hubble parameter in such theories grows along the Planck boundary.
Therefore the leading contribution to the volume of the Universe is given by
the domains expanding with ever growing speed and containing indefinitely large
values of the fields $\sigma$ and $\phi$. The diffusion  equations describing
this process have runaway solutions which describe
distributions $P_p(\phi,\sigma,t)$ running towards infinitely large $\sigma$.
In many cases, the center  of the distribution $P_p$ reaches infinitely large
values
of $\sigma $ within finite time.

This does not necessarily mean that the corresponding theories are physically
unacceptable. It may happen that our idea that we should live in a part of the
Universe corresponding to a maximum of  $P_p(\phi,\sigma,t)$ at a given time
$t$ is incorrect; see Ref.~\cite{GBLL} for a discussion of this issue. In
particular, as it was shown in  \cite{LLM,GBLL}, the behavior of the
distribution $P_p(\phi,\sigma,t)$ depends on the
choice of time parametrization.  If one studies, e.g. the
probability distribution $P_p(\phi,\sigma,\tau)$, where $\tau =
\ln a(t)$, its behavior may be quite different,
and it may not exhibit any runaway solutions.  Moreover, it is
very easy to stabilize the distribution $P_p(\phi,\sigma,t)$ by
making the effective potential $V(\sigma)$ very curved and
unsuitable for inflation at large $\sigma$. Still it would be
interesting to study this phenomenon in a more detailed way and
to see whether the  distribution $P_p(\phi,\sigma,t)$ in the
Brans--Dicke theory  can be stabilized in a natural way. If this
stabilization can occur only at very large values of the field
$\sigma$, then
the typical value of the Brans--Dicke field $\phi$ at the end of
inflation will also be extremely large. This may give us
a tentative explanation of the anomalously large
value of the Planck mass.

\section{\label{3} A more general theory}

Let us now turn to a  scenario in which some of the
problem discussed above could be resolved. For that purpose
we will consider the classical evolution of the inflaton field
with a generic chaotic potential, in the context of the
Jordan--Brans--Dicke theory of gravity and a curvature
coupled inflaton with a non minimal coupling $\xi$,
\begin{equation}\label{S}
{\cal S}=\int d^4x \sqrt{-g} \left[{1\over8\omega} \phi^2 R -
{1\over2} \xi \sigma^2 R - {1\over2}(\partial\phi)^2 -
{1\over2}(\partial\sigma)^2 - V(\sigma)\right]\ .
\end{equation}
In this theory the Planck mass takes the form
\begin{equation}\label{PM}
M_{\rm P}^2(\phi,\sigma) = 16\pi\Phi = 8\pi \left({1\over4\omega}
\phi^2 - \xi\sigma^2\right)\ ,
\end{equation}
where $\Phi$ plays the role of the Brans--Dicke scalar.
The value of the Brans--Dicke parameter $\omega$ is bounded by
the post-Newtonian experiments~\cite{TEGP} and primordial
nucleosynthesis~\cite{PNB} to be very large, $\omega > 500$, and
therefore it is reasonable to use the approximation $\omega \gg
1$ in the following analysis. The parameter $\xi$, on the other
hand, is unconstrained and could take positive or negative
values. Here we will consider the case of a small and {\em negative}
$\xi$.

The equations of motion for the theory (\ref{S})
can be written as
\begin{equation}\begin{array}{rl}\label{XEQ}
\nabla^2\phi\ = &{\displaystyle\!{1\over4\omega} \phi R\ ,}\\[3mm]
\nabla^2\sigma\ = &{\displaystyle\!-\,V'(\sigma) -
\xi\sigma R\ , }
\end{array}\end{equation}
\begin{equation}\begin{array}{rl}
{\displaystyle
-\left({1\over4\omega}\phi^2 - \xi\sigma^2\right)\left(R_{\mu\nu}
- {1\over2} g_{\mu\nu} R\right)}
&{=\ \displaystyle\! g_{\mu\nu} V(\sigma) +
\left(\nabla_\mu\nabla_\nu - g_{\mu\nu} \nabla^2\right)
\left({1\over4\omega}\phi^2 - \xi\sigma^2\right) }\\[3mm]\nonumber
&{\displaystyle + \left(\partial_\mu\phi\partial_\nu\phi -
{1\over2} g_{\mu\nu} (\partial\phi)^2\right) +
\left(\partial_\mu\sigma\partial_\nu\sigma -
{1\over2} g_{\mu\nu} (\partial\sigma)^2\right) }\ .
\end{array}\end{equation}\nonumber
We can then write the exact equations for the homogeneous fields
in flat space ($k=0$) as
\begin{equation}\label{REL}
{1\over2}\left[\Big(1+{3\over2\omega}\Big)\,\nabla^2\phi^2 +
(1 - 6\xi)\,\nabla^2\sigma^2 \right] =
4 V(\sigma) - \sigma V'(\sigma)\ ,
\end{equation}
\begin{equation}\begin{array}{rl}\label{SEQ2}
{\displaystyle \left[{1\over4\omega}\Big(1+{3\over2\omega}\Big)
\phi^2 - \xi(1 - 6\xi)\sigma^2\right] R }\ =
&{\displaystyle\! 4V(\sigma) - 6\xi\,\sigma V'(\sigma)- (1 - 6\xi)
\dot\sigma^2 - \Big(1+{3\over2\omega}\Big)\dot\phi^2}\ ,\\[4mm]
{\displaystyle \left({1\over4\omega}\phi^2 - \xi\sigma^2\right)\
3H^2}\ = &{\displaystyle\! V(\sigma) + 6\xi\, H\dot\sigma\sigma +
{1\over2} \dot\sigma^2 - {3\over2\omega}\, H\dot\phi\phi +
{1\over2} \dot\phi^2 } \ ,\vspace{3mm}
\end{array}\end{equation}
where $R = 12 H^2 + 6 \dot H$. Note that for $\xi = 0$ and
$\Phi = \phi^2/8\omega$ we recover the usual BD equations. From now
on we will drop the last two terms of Eqs.~(\ref{SEQ2}) since
they are subleading for large $\omega$.

During inflation, we can write the equations of motion of the
homogeneous fields $\phi$ and $\sigma$ for small $|\xi|$ and
large $\omega$, in the slow-roll approximation $(\ddot\phi \ll
H\dot\phi \ll H^2\phi)$, as
\begin{equation}\begin{array}{rl}\label{SEQ}
\dot\phi\ = &{\displaystyle\!{H\phi\over\omega}\ ,}\\[4mm]
\dot\sigma\ = &{\displaystyle\!{H\sigma\over\omega}\left[
\left(1 - {\sigma V'\over4V}\right)\left({\phi^2\over\sigma^2}
+ \mu\right) - {\phi^2\over\sigma^2}\right] \ ,}\\[4mm]
H^2\ = &{\displaystyle\!{\omega\over3}\
{4V(\sigma)\over\phi^2 + \mu\,\sigma^2} }\ ,
\end{array}\end{equation}
where $\mu \equiv 4\omega|\xi|$. Note that for small $\mu$
they reduce to the usual BD equations.

\subsection{\label{4}Massive curvature-coupled inflaton}

Let us consider in some detail the theory (\ref{S}) with
$V(\sigma) = {1\over2} m^2\sigma^2$ at large $\sigma$ and very
small $\xi < 0$.  In this case it is useful to make a change of
variables to polar coordinates $r^2 = \phi^2 + \varphi^2, \ z =
{\phi\over\varphi},\ \varphi^2 = 2(1+6|\xi|)\sigma^2 \simeq
2\sigma^2$. In these coordinates, the
equations of motion (\ref{SEQ}) take the form
\begin{equation}\begin{array}{rl}\label{EQS}
&{\displaystyle \dot z={z H\over\omega}\Big(z^2 + 1 - \mu/2
\Big)\ ,}\\[3mm]
&{\displaystyle \dot r= {r H\over\omega}\,
{\mu/2\over1+z^2}\ ,}\\[3mm]
&{\displaystyle H^2 = {\omega\over3}\ {m^2\over z^2+\mu/2}\ . }
\end{array}\end{equation}
One can solve these equations parametrically in the
$z$-variable, for $\mu \neq 2$,
\begin{equation}\begin{array}{rl}\label{DRZ}
&{\displaystyle {dr\over dz} = {r\,\mu/2\over z(1+z^2)
(z^2 + 1 - \mu/2)}\ , \hspace{1.5cm} r_o = r(z_o)\ ,}\\[5mm]
&{\displaystyle r(z) = r_o \left({z\over z_o}
\right)^{\mu/(2-\mu)}\ \left({1+z^2\over1+z_o^2}
\right)^{1/2}\ \left({z^2+1-\mu/2\over z_o^2+1-\mu/2}
\right)^{-1/(2-\mu)}\ ,}\\[3mm]
\end{array}\end{equation}
which gives
\begin{equation}\begin{array}{rl}\label{PSZ}
&{\displaystyle \phi(z)=r(z)\ (1+z^{-2})^{-1/2}\ ,}\\[3mm]
&{\displaystyle \varphi(z)=r(z)\ (1+z^2)^{-1/2}\ . }
\end{array}\end{equation}
For $|\xi| \ll 1$, the condition for existence of inflation
($|\dot H| < H^2$) is $\ \dot\sigma^2 < V$,
\begin{equation}\label{REI}
(z^2 - \mu/2)^2 < {3\omega\over2}\,(z^2 + \mu/2)\ ,
\end{equation}
which gives a slightly modified condition for the end of
inflation, $z_e^2 + \mu/2 = 3\omega/2$.

For very small $\mu$, the trajectory
is very close to a circle in the plane $(\varphi,\phi)$.
However, for arbitrary $\mu$ the scalar fields move along a more
complicated trajectory. There are regions
of parameter space for which inflation never ends. For instance,
if $\mu > 2$ and $z^2 < (\mu - 2)/2$, then $\dot z < 0$ and inflation
never ends (unless the field $\sigma$ tunnels (diffuses) to its large
values quantum mechanically). This is a rather surprising result since
there is nothing singular in our equations (\ref{EQS}) at $\mu = 2$.
The reason of this unexpected behavior is that the field $\sigma$ in
the regime $\phi \ll \sigma$ for $\mu > 2$ grows much faster than the
field $\phi$, so it never approaches the end of inflation with $\phi
\gg \sigma$.

The value of the effective Planck mass at the end of inflation
is given by (see Eq.\ref{DRZ}),
\begin{equation}\label{MPEND}
M_{\rm P}  = \sqrt{2\pi\over\omega}\ r(z_e)\
\left({z_e^2+\mu/2\over z_e^2+1}\right)^{1/2}\ \simeq\
\sqrt{2\pi\over\omega}\ r_o\,z_o\,
\left({2-\mu\over2z_o^2}\right)^{1\over2-\mu}\ ,
\end{equation}
where we have used the fact that $z_e\gg1$ and $z_o\ll1$. Note that
the value of the Planck mass at the end of inflation is very sensitive to
initial conditions
and, for the reason explained above, it becomes exponentially large
when $\mu \to 2$: \
$\ M_{\rm P}  = \sqrt{4\pi|\xi|}\ \phi_o\,
\exp\Big(z_o^{-2}\Big)$.

We should now discuss possible initial conditions and the regime of
self-reproduction of the Universe in this model. The Planck boundary
in this theory looks like an ellipse
\begin{equation}\begin{array}{c}\label{SIG}
{\displaystyle \phi^2 + \mu \left(\sigma-{\sigma_{\rm max}\over2}
\right)^2 = {\mu\sigma_{\rm max}^2\over4}\ ,}\\[4mm]
{\displaystyle \sigma_{\rm max} \equiv {m\over8\pi\sqrt{2}|\xi|}\ .}
\end{array}\end{equation}
There is no constraint on $\mu$; however, we will assume that $\xi$
and $1/\omega$ are both of the same order of magnitude. We will see
that this is a consistent approximation in this scenario.

The amplitude of quantum fluctuations of the scalar fields,
$\delta\phi$ and $\delta\sigma$, whose wavelengths are stretched
beyond the horizon and act on the quasi--homogeneous background
fields like a  stochastic force, can be computed as in
\cite{GBLL,JGB} and they are given by (for $\omega \gg 1$ and
$|\xi| \ll 1$)
\begin{equation}\label{STEP}
\delta\phi \simeq {H\over2\pi}\ ,\hspace{2cm}
\delta\sigma \simeq {H\over2\pi}\ .
\end{equation}
Density perturbations in our model can be calculated  as in Refs.
\cite{GBLL,JGB,STAR}. For the theory $V(\sigma) = {1\over2} m^2
\sigma^2$, in the large $\omega$ limit\footnote{Note that during
the last stages of inflation, there is an approximate equivalence
of the Einstein and Jordan frames.}, we find
\begin{equation}\label{RPE}
{\delta\rho\over\rho} \sim   {50 \,m\over M_{\rm P}} \ .
\end{equation}
Note that the larger is the Planck mass at the end of inflation
in a given region of the Universe, the smaller will be  density
perturbations in this region. This suggests that the large value
of Planck mass $M_{\rm P}$   {\it in our part of the Universe} may
be related to the small value of the amplitude of density perturbations
${\delta\rho\over\rho} \sim   5\cdot 10^{-5}$ \cite{GBLL}.
In other words, instead of two independent small parameters,
${m\over M_{\rm P}} \ll 1$ and ${\delta\rho\over\rho} \sim
5\cdot 10^{-5}$, we have only one.

As we already discussed in the previous section,  those inflationary
domains where quantum jumps during the time $H^{-1}$ are more
important than their classical motion, enter the regime of self
reproduction. One can easily show that in our case the region where
this regime is possible is the interior of the ellipse
\begin{equation}\begin{array}{c}\label{BIP}
{\displaystyle \phi^2 + \nu\left(\sigma -
{\sigma_\ast\over2}\right)^2 = {\nu\sigma_\ast^2\over4}
\ ,}\\[4mm]
{\displaystyle \sigma_\ast \equiv
{m\over4\pi\sqrt6|\xi|^{3/2}}, \hspace{1cm}
\nu \equiv \mu - 1 + (\mu^2-\mu+1)^{1/2}\ .}
\end{array}\end{equation}

In our model the Hubble constant takes its maximal values along the
line $\phi=0$ (i.e. $z = 0$). Along this line  there is a wide
region where self-reproduction is possible. This region begins at the
Planck boundary (\ref{SIG}) and ends at the self-reproduction boundary
(\ref{BIP}):
\begin{equation}\label{SIB}
{m\over8\pi\sqrt{2}|\xi|} < \sigma < {m\over4\pi\sqrt6
|\xi|^{3/2}}\ .
\end{equation}
Since this interval is limited, runaway solutions here are impossible. In such
a situation the probability distribution $P_p(\phi,\sigma,t)$ should be
stationary, with a  maximum somewhere at this line, in the interval
(\ref{SIB}).

Note, however, that this maximum should be relatively smooth, since the
value of the Hubble constant does not depend on $\sigma$ along the line
$\phi = 0$:
\begin{equation}\label{DPHI}
{H(\phi=0) } = {m\over \sqrt{6}|\xi|^{1/2}}\ .
\end{equation}
For a complete investigation of the probability distribution
$P_p(\phi,\sigma,t)$ and of the   value of $M_{\rm P}$ at the end of
inflation one should solve numerically the
diffusion equations for $P_p(\phi,\sigma,t)$,
as we did in Ref.~\cite{GBLL}. To get a rough idea of the resulting
distribution of possible values of $M_{\rm P}$
at the end of inflation one can simply take all points at the boundary of
self-reproduction and treat
them as initial conditions for the solutions (\ref{DRZ}), (\ref{MPEND}).
However, one can get a much better picture if one takes into account that the
main contribution to the probability distribution on this boundary is given by
the part of the boundary of the region of self-reproduction with the highest
value of $H$. This value is achieved for $\sigma =  \sigma_\ast\ $
(\ref{BIP}) at the line $\phi = 0$.
The peak cannot be wider than ${\sqrt\nu\over 2}\sigma_\ast$
in the $\phi$-direction, and in fact it is expected to be much more narrow.
Estimates of the values of the effective Planck mass generated
after inflation suggest that $M_{\rm P}$  can take all possible values
from $0$ to $\infty$, but typical values are much greater than
${m\over\sqrt\mu\xi}$. If, for example, one considers the evolution
of domains with initial values of the fields   $\sigma \sim  \sigma_\ast$
and $\phi
{\ \lower-1.2pt\vbox{\hbox{\rlap{$>$}\lower5pt\vbox{\hbox{$\sim$}}}}\ }
{H\over2\pi}$ (due to unavoidable quantum jumps near $\phi = 0$), Eq.
(\ref{MPEND}) yields
\begin{equation}\label{MPEFF}
{m\over M_{\rm P}}
{\ \lower-1.2pt\vbox{\hbox{\rlap{$>$}\lower5pt\vbox{\hbox{$\sim$}}}}\ }
\sqrt{3\pi\mu} \left({4|\xi|^2\over2-\mu}\right)^{1\over2-\mu}\ .
\end{equation}

One can then estimate the allowed values of $|\xi|$ from the
amplitude of density perturbations (\ref{RPE}). For $\mu\simeq1$ we
find $|\xi|
{\ \lower-1.2pt\vbox{\hbox{\rlap{$<$}\lower5pt\vbox{\hbox{$\sim$}}}}\ }
4\cdot 10^{-4}$, which is consistent with the constraint  $\omega>500$
and our condition $\mu\simeq1$.
Note also that ${m\over M_{\rm P}} \simeq \sqrt{6\pi}\,\exp
\Big(-{1\over 2|\xi|^2}\Big)$ for $\mu \to 2$,and it becomes very
easy to obtain a very small ratio of ${m\over M_{\rm P}}$ even for
not too small values of $|\xi|$. Therefore, we have a natural
realization of our model without the need for very small numbers.
In fact one could argue that $\omega\sim500$ is not such a large
number, if we understand it as the dimensionless coupling
$\beta$ of matter to the dilaton in the Einstein frame
\cite{DIL}, $2\beta=(2\omega+3)^{-1/2} \sim 0.03$.

\subsection{Self-coupled inflaton}

We must now consider possible corrections to the inflaton potential,
in order to see whether our results are stable with respect to small
modifications of the theory. Let us add an inflaton self-coupling to the
theory,
$V(\sigma) = {1\over2} m^2 \sigma^2 + {\lambda\over4} \sigma^4$.
The corresponding equations of motion read
\begin{equation}\begin{array}{rl}\label{EQSP}
&{\displaystyle \dot z={z H\over\omega} \, {2m^2(1+z^2)(2z^2 + 2 - {\mu})+
{\lambda r^2(1+2z^2)}\over 4m^2(1+z^2)+{\lambda r^2 }}
 \ ,}\\[3mm]
&{\displaystyle \dot r= {r H\over\omega}\, {2m^2(1+z^2)\mu-\lambda r^2
z^2\over(4m^2(1+z^2)+\lambda r^2)(1+z^2)} \ ,}\\[3mm]
&{\displaystyle H^2 = {\omega\over6}\, {4m^2
(1+z^2)+\lambda r^2\over (1+z^2)(2z^2+\mu)} \ . }
\end{array}\end{equation}

There is a bifurcation line in the plane $(z,r)$ that separates
domains for which inflation never ends. This is the line $\dot z
= 0$,
\begin{equation}\label{ZET}
z^2 = {\mu-2\over2} - {\lambda r^2\over4m^2}\,{1+2z^2\over1+z^2}\ .
\end{equation}
For $\mu > 2$ and $r^2 < 2m^2(\mu-2)/\lambda$, there are values
of $z$ for which $\dot z< 0$, which could suggest, just as in the previous
model  with $V(\sigma) = {1\over2} m^2 \sigma^2$, that the end of inflation
cannot be reached.
However, as $r$ increases (\ref{EQSP}), $\dot z$ eventually
changes sign and  inflation ends.

The boundary of self-reproduction in this theory is given by
\begin{equation}\label{SRL}
\Big(\phi^2 + \nu\sigma^2\Big)\left(1 - {\phi^2\over\sigma_c^2}
\right) = \nu\sigma_\ast\sigma\,
\left(1 + {\sigma^2\over\sigma_c^2}\right)^{1/2}\ ,
\end{equation}
where $\nu$ and $\sigma_\ast$ were defined in (\ref{BIP}).

The Planck boundary for this theory is
\begin{equation}\label{PBE}
\phi^2 + \mu\sigma^2 = \mu\sigma_{\rm max}\sigma\,
\left(1 + {\sigma^2\over\sigma_c^2}\right)^{1/2}\ ,
\end{equation}
where $\sigma_{\rm max}$ was given in (\ref{SIG}) and $\sigma_c \equiv
2m^2/\sqrt\lambda$. At small $\sigma$ this boundary looks like an ellipse, but
at large $\sigma$ it becomes a straight line.  For $\sigma_{\rm max} <
\sigma_c$  these
two parts of the Planck boundary are disconnected: both lines  cross the
$\sigma$ axis at different points. For $\sigma_{\rm max} > \sigma_c$ these two
lines
form a continuous curve.  The greatest values of the Hubble parameter appear
along the part of the Planck boundary with $\sigma \to \infty$. Therefore in
this model there  always exist runaway solutions describing the probability
distribution $P_p(\phi,\sigma,t)$ rapidly moving towards indefinitely large
values of $\sigma$ and $\phi$.
In other words, the existence of a stationary regime which we have found in the
previous model appears to be unstable with respect to a small modification of
the effective potential of the inflaton field. It makes this model less
attractive.

\section{\label{4a}One loop corrections}

Now let us return back to the simple Brans--Dicke theory (\ref{SP}) with the
inflaton field minimally coupled to gravity, but let us take into account
quantum corrections to the effective potential $V(\sigma)$. In the one-loop
approximation one can represent the effective potential in the following way:
\begin{equation}\label{LOOP2}
V(\sigma) = {\lambda\over4} (\sigma^2 - \sigma_0^2)^2 +
\beta \sigma^4 \ln{\sigma\over\sigma_0}\ ,
\end{equation}
where $\sigma_0^2 = m^2/\lambda$ is the value of the inflaton
field at the minimum of the potential, and $\beta$ depends on
the values of coupling constants in the theory. For example, if the scalar
field $\sigma$ interacts with the fermion field with the coupling constant $h$,
one obtains $\beta = \left(9\lambda^2 - 4h^2\right)/32\pi^2$~\cite{KriveLinde}.
Note, that $\beta$ will be negative for $h^2 \gg \lambda^2$.
In that case, a new maximum
will appear at exponentially large $\sigma$,
\begin{equation}\label{MAX}
\sigma_{\rm max} \simeq \sigma_0\, \exp{\lambda\over4|\beta|} \gg
\sigma_0\ ,
\end{equation}
followed by a very sharp fall-off to negative values of the
potential, rendering the vacuum unstable~\cite{KriveLinde}. In our case
this sharp fall-off may act effectively as a natural boundary
for the diffusion of the inflaton field $\sigma$. In such a situation the
probability distribution $P_p(\phi,\sigma,t)$ becomes stationary, with the
maximum concentrated near the Planck boundary at $\sigma = \sigma_{\rm max}$.

In the theory (\ref{LOOP2}), the   equations of motion
for $\sigma\gg\sigma_0$ and $\xi=0$, become
\begin{equation}\begin{array}{rl}\label{QSE}
&{\displaystyle \dot\phi = {H\phi\over\omega}\ ,}\\[3mm]
&{\displaystyle \dot\sigma = - {V'(\sigma)\over3H} \simeq
- {H\phi^2\over\omega\sigma}\ ,}\\[3mm]
&{\displaystyle H^2 = {4\omega V(\sigma)\over3\phi^2} }\ .
\end{array}\end{equation}
For $\sigma_0 \ll \sigma \ll \sigma_{\rm max}$ the classical motion is
circular, just as in the theory $\lambda\sigma^4$, and we can define polar
coordinates $(r,z)$ as usual, with $z=\phi/\sigma$. Therefore one can obtain
the following estimate of the Planck mass after the end of inflation:
\begin{equation}\label{PlankMass}
M_{\rm P} \sim \sqrt{2\pi\over\omega}\,\sigma_{\rm max}
\sim \sqrt{2\pi\over \omega}\,\sigma_0\, \exp{\lambda\over4|\beta|} \ .
\end{equation}
Note that in this case the effective Planck mass for
$|\beta| \ll \lambda$ becomes exponentially large. In the regime when it is
much greater than $\sigma_0$ the amplitude of density perturbations is
given by the standard expression ${\delta\rho\over \rho} \sim 10^2
\sqrt\lambda$. Thus, one does not get anything new from the point of view of
density perturbations, but one can obtain a natural explanation of the very
large value of the Planck mass as compared with other mass scales in the
theory.

\section{\label{5}Non-perturbative effects}

All results obtained in the present paper are related to the probability
distribution $P_p(\phi,\sigma,t)$, which shows the fraction of the volume of
the Universe with the fields $\phi$ and $\sigma$ at a given moment of time $t$
in synchronous coordinates.
We should emphasize again that this is not a unique choice of measure in
quantum cosmology. One could study, for example, the probability distribution
$P_p(\phi,\sigma,\tau)$, where $\tau \sim \ln a(t)$. This distribution is
considerably different from $P_p(\phi,\sigma,t)$. This does not allow us to
make unambiguous predictions until the issue of measure in quantum cosmology is
resolved, see a discussion of this issue in Refs.~\cite{LLM,GBLL,GBL}.
Still we believe that investigation of
$P_p(\phi,\sigma,t)$ gives us  very interesting information about the
structure of inflationary universe.

Recently  it was shown, in the context of chaotic inflation based on the usual
Einstein theory,  that   the main fraction of
volume of the  Universe in a state with a given density $\rho$ at any given
moment of time $t$ is concentrated near the centers
of deep exponentially wide spherically symmetric holes in the density
distribution~\cite{OPEN}.   For the reason discussed above, interpretation of
this result is not unambiguous, and we are not sure that it implies that we
must live near the center of a spherically symmetric void. However, we think
that this result is very nontrivial and deserves further investigation
\cite{OPEN2}.

Here we would like to study whether a similar effect occurs in the context of
the Brans--Dicke theory. As we will see, the effect does take place, but its
amplitude is significantly different. In order to understand it, we should
briefly remind the origin of this effect in the Einstein theory, with one
scalar field $\sigma$ (the inflation field), and then we will make a
generalization to the Brans--Dicke case.

The best way  to examine this scenario is to investigate the probability
distribution $P_p(\sigma,t)$.
The distribution $P_p(\sigma,t)$ obeys the following diffusion
equation (see Ref.~\cite{LLM} and references therein):
\begin{equation}\label{E372}
\frac{\partial P_p}{\partial t} =
 \frac{1}{2}  \frac{\partial }{\partial\sigma}
 \left( \frac{H^{3/2}(\sigma)}{2\pi}\, \frac{\partial }{\partial\sigma}
\Bigl(
 \frac{H^{3/2}(\sigma)}{2\pi}  P_p \Bigr) +
\frac{V'(\sigma)}{3H(\sigma)} \, P_p\right)  +  3H(\sigma)  P_p\ .
\end{equation}
Here we temporarily use the system of units $M_{\rm P} = 1$. One may try
to obtain solutions of equation (\ref{E372}) in the
form of the  series $P_p(\sigma,t) =
\sum_{s=1}^{\infty} { e^{\lambda_s t}\,   \pi_s(\sigma) } $. In the limit of
large time $t$ only the term with the largest eigenvalue $\lambda_1$ survives,
$P_p(\sigma,t) =
e^{\lambda_1 t}\, \pi_1(\sigma) $. The function $\pi_1$ in the limit $t \to
\infty$ has a meaning of a normalized   time-independent  probability
distribution to find a given field $\sigma$ in a unit physical volume, whereas
the function  $e^{\lambda_1 t}$ shows the overall growth of the volume of all
parts of the Universe, which does  not depend on $\sigma$ in the limit $t\to
\infty$. In this limit  one can write Eq.~(\ref{E372}) in the form
\begin{equation} \label{eq17}
\frac{1}{2}  \frac{\partial }{\partial\sigma}
 \left( \frac{H^{3/2}(\sigma)}{2\pi} \frac{\partial }{\partial\sigma}
\left(
 \frac{H^{3/2}(\sigma)}{2\pi} \pi_1(\sigma) \right) \right)
 + \frac{\partial }{\partial\sigma} \left( \frac{V'(\sigma)}{3H(\sigma)} \,
\pi_1(\sigma) \right)
 + 3H(\sigma)  \cdot \pi_1(\sigma) =
\lambda_1 \, \pi_1(\sigma)  \ .
\end{equation}
In the simplest theory with $V(\sigma) = {\lambda\over 4} \sigma^4\,$ and
$\ H = \sqrt{2\pi\lambda \over 3}\sigma^2$, Eq.~(\ref{eq17}) reads
\cite{LLM}:
\begin{equation}\label{r1}
\pi_1'' + \pi_1'\Bigl({6\over\lambda\sigma^5} +
{9\over\sigma}\Bigr) + \pi_1 \Bigl({6\over\lambda\sigma^6} +
{15\over\sigma^2} +
{36\pi\over\lambda\sigma^4} -{\lambda_1\over \pi \sigma^6}
\Bigl({6\pi\over\lambda}\Bigr)^{3/2}\Bigr) = 0\ .
\end{equation}
This equation can be solved both analytically and numerically. The result is
that the eigenvalue $\lambda_1$ is given by $d(\lambda)\, H_{\rm max}$.
Here $d(\lambda)$ is the fractal dimension, which approaches $3$ in the limit
$\lambda \to 0$, while $H_{\rm max}$  is the maximum possible value
of the Hubble constant during inflation, which in our case corresponds to its
value at the
Planck boundary $V(\sigma) = M_{\rm P}^4= 1$. This gives  $H_{\rm max}  =
{2\sqrt{2\pi
\over 3}}$. Thus, in the small
$\lambda$ limit one has $\lambda_1 = 3 H_{\rm max}
= 2\sqrt{6\pi} \approx 8.68$.

The distribution $\pi_1$ depends on $\sigma$   very sharply. One can easily
check that   at small $\sigma$ the leading terms in
Eq.~(\ref{r1}) are the second and the last ones.
(This means, in particular, that the diffusion terms in
Eqs.~(\ref{E372}),(\ref{eq17}) can be neglected.) Therefore the
solution of equation (\ref{r1}) for small $\sigma$  corresponding to the last
stages of inflation is given by
\begin{equation}\label{SMALLPHI}
\pi_1 \sim     \sigma^{\sqrt{6\pi\over
\lambda}\lambda_1} \,   \  .
\end{equation}
This is an extremely strong dependence. For example, $\pi_1 \sim
\sigma^{10^8}$ for the realistic value $\lambda \sim 10^{-13}$.

Consider   all inflationary domains which contain a given field $\sigma$
at a given moment of time $t$. Let us try to find a typical value of
this field in those domains at the earlier moment $t - H^{-1}$.
In order to do it, one should add to
$\sigma$ the value of its classical drift $\dot\sigma H^{-1}$.
One should also add the
amplitude of  quantum jumps $\Delta \sigma$. The usual estimate of the
magnitude of a typical jump is
$\pm {H\over 2\pi}$.  This is a correct estimate if we are interested in a
typical amplitude of jumps at any given point.  However, if we are considering
{\it all} domains with a given $\sigma$ and trying to find {\it all} those
domains from which
the field $\sigma$ could originate, the answer for may be quite
different. The total volume of all domains with a given field $\sigma$
at any moment of time $t$ strongly depends on $\sigma$:
\,${P}_p(\sigma) \sim \pi_1(\sigma)
\sim   \sigma^{\sqrt{6\pi\over \lambda}\lambda_1}$, see Eq.~(\ref{SMALLPHI}).
This means that the total volume of all domains which could
jump towards the given field $\sigma$ from the value $\sigma +\Delta \sigma$
will be
enhanced by a large  additional factor $ { {P}_p(\sigma +\Delta \sigma)\over
 {P}_p(\sigma)} \sim   \Bigl(1+{\Delta\sigma\over
\sigma}\Bigr)^{\sqrt{6\pi\over
\lambda}\lambda_1}$. On the other hand, the probability of large jumps
$\Delta\sigma$ is suppressed by the Gaussian factor
$\exp\Bigl(-{2\pi^2\Delta\sigma^2\over H^2}\Bigr)$.
The product of these two factors has a sharp maximum
at $\Delta\sigma = \lambda_1 \sigma   \cdot {H\over 2\pi}$. In other words,
most of the domains of
a given field $\sigma$ are formed due to the   jumps which have definite sign
(they are decreasing the value of the scalar field), and which  are greater
than the
``typical'' ones by the  amplification factor  $N = \lambda_1 \sigma $. In
the usual notation, this amplification factor is given by
\begin{equation}\label{x1}
N = {\lambda_1 \sigma\over M^2_{\rm P}} \ ,
\end{equation}
where $\lambda_1 = 3 H_{\rm max}
= 2\sqrt{6\pi} M_{\rm P} \approx 8.68 M_{\rm P}$.
For $\sigma \sim 4.5 M_{\rm P}$ (the scale at
which the large scale structure of our part of the Universe has been formed),
the amplification factor in our theory is about $  40$. However, the   value
of this factor is very sensitive to our assumptions concerning the Planck
boundary, it can be much bigger or much smaller than $40$. As explained in
\cite{OPEN}, this effect does not alter the standard theory of density
perturbations in inflationary universe, but it puts these perturbations on the
top (or, more precisely, to the bottom) of the   distribution of the scalar
field $\sigma$ which appears as a result of  its large jumps.

We can now return to the Brans--Dicke theory with $\omega \gg 1$. In this case
both fields $\phi$ and $\sigma$ move and fluctuate, and therefore one should
write a two-dimensional diffusion equation for these fields~\cite{GBLL}.
 However, at the last stages of inflation the classical motion of the scalar
field $\phi$ is very slow, and, as we argued above, diffusion in the first
approximation can be neglected. Then the problem reduces to the one we have
already solved, and the amplification coefficient  will be given by eq.
(\ref{x1}). The only difference is that now the value of the coefficient
$\lambda_1 = 3 H_{\rm max}$ will be determined not by  the Planck mass
at the end of inflation, but by the much smaller Planck mass at the place near
the Planck boundary corresponding to the peak of the distribution
$P_p(\phi,\sigma)$. If the distribution is stationary due to the existence of
some kind of boundary at $\sigma = \sigma_{\rm max}$, then  the peak is
concentrated at the Planck boundary near $\sigma_{\rm max}$. In this case one
can show that  $\lambda_1 = 3 H(\sigma_{\rm max}) = 2 \sigma_{\rm max}
\sqrt{3\pi \sqrt {\lambda}}$, for $V(\sigma) = {\lambda\over4}\,\sigma^4$.
Meanwhile, the typical Planck mass after
inflation is given by $M_{\rm P} \sim \sqrt{2\pi\over\omega}\sigma_{\rm max}$
\cite{GBLL}. This leads to the following realization of Eq.~(\ref{x1}) for the
Brans--Dicke theory:
\begin{equation}\label{x2}
N \sim \left(6\omega\sqrt\lambda\right)^{1/2}\,{\sigma\over M_{\rm P}} \ ,
\end{equation}
which gives $N \sim 4.5  \sqrt{6\omega\sqrt\lambda}$,
for $\sigma \sim 4.5 M_{\rm P}$.
Note that for $\lambda \sim 10^{-13}$ and $\omega \sim 10^{3}$ the factor $N$
becomes very small, and therefore all the nonperturbative effects
discussed above will be negligible, contrary to the case in general
relativity~\cite{OPEN}.

Note that in this investigation we have assumed the existence of the
upper boundary $\sigma_{\rm max}$. The situation will be quite different if
there were no stationary solutions for $P_p$. In such a case the
nonperturbative effects will be extremely strong. However, then we will meet
the problem of runaway solutions which would suggest that
$M_{\rm P} \to \infty$ in
the main part of the physical volume of the Universe. In this paper we have
shown that in principle it is possible to avoid runaway solutions in the
Brans--Dicke theory. But then from our results it follows that the same trick
which makes the distribution stationary simultaneously kills the
nonperturbative effects in the inflationary Brans--Dicke cosmology.

\section{\label{6}Conclusions}

Investigation of the probability distribution $P_p$ gives us a lot of
interesting information about the properties of the inflationary universe.
Some of
these properties (such as the very existence of the regime of self-reproduction
and the fractal structure of the Universe) do not depend on the choice of time
parametrization, and therefore their interpretation is relatively
straightforward. Some other properties of $P_p$ do depend on the choice of time
parametrization. Sometimes it is not enough   to know  $P_p$, we need to know
also whether all parts of the Universe which are described by this distribution
are equally well suited for existence of life. In such situations
interpretation of the results becomes increasingly speculative. Nevertheless we
believe that even in these cases investigation of $P_p$ and attempts of its
interpretation can be very useful. We are  learning how to formulate questions
in the context of quantum cosmology.  In some cases we are obtaining results
which look obviously incorrect or contradict observational data; then we may
conclude that we are using quantum cosmology in a wrong way. In some other
cases quantum cosmology allows us to obtain important results which cannot be
obtained by other methods. This may be considered as an indication that we are
on the right track.  Thus, by this   trial and error method we may finally
learn how to use quantum  cosmology.

To give a particular example one may consider the old issue of the wave
function of the Universe. The two most popular candidates are the
Hartle-Hawking wave function, which in the context of inflationary cosmology
reads $\exp\Bigl(\frac{3 M_{\rm P}^4}{16V(\phi)}\Bigr)$~\cite{HH}, and
the tunneling wave function
$\exp\Bigl(-\frac{3 M_{\rm P}^4}{16V(\phi)}\Bigr)$~\cite{Creation}.
Neither of these two functions was rigorously derived. For some reason which is
not related to its derivation, the square of the Hartle-Hawking wave function
correctly describes tunneling between two different de Sitter universes with
different values of $V(\phi)$~\cite{Tunn}. However if one makes an attempt to
apply it to the probability of creation of the inflationary universe, one comes
to a physically incorrect conclusion that it is much easier to create an
infinitely large inflationary universe with $V(\phi) \to 0$ than a Planck-size
universe with $V(\phi) \sim M_{\rm P}^4$. The tunneling wave function leads
to a qualitatively correct description of quantum creation of the Universe
``from nothing,''  but one should not uncritically  apply it to, e.g. the
formation of black holes~\cite{HR}. The reasons for the
limited applicability of each of these functions are explained
in Ref.~\cite{Book}, whereas in Ref.~\cite{LLM} it is shown that
expressions of the type of $\exp\Bigl(\frac{3 M_{\rm P}^4}{16V(\phi)}\Bigr)$ or
$\exp\Bigl(-\frac{3 M_{\rm P}^4}{16V(\phi)}\Bigr)$ appear in many problems of
quantum cosmology which are not related in any obvious way to the original
``derivations'' of the Hartle-Hawking and tunneling wave functions.

Something similar may occur with our investigation of the probability
distribution $P_p$. This distribution is certainly very useful, but sometimes
it becomes   tempting to use it in the situations where its interpretation
is ambiguous and the final success is not   guaranteed. However, the
possibility to obtain very strong results and to look at the interplay between
particle physics and cosmology from an entirely new point of view suggests us
to continue this trial-and-error investigation. Here we may mention our attempt
to address the cosmological constant problem in the context of the Starobinsky
model~\cite{GBL} and the possibility to explain why in the main part of the
Universe the scalar-tensor theories of gravity are reduced to the Einstein
theory~\cite{JGBDW}. In this paper (see also our previous publications
\cite{ExtChaot,GBLL,JGB}) we demonstrated that in the context of inflationary
Brans--Dicke theory it may be  the possible to explain the anomalously large
value of the Planck mass $M_{\rm P}$. We have
shown also that the structure of our part of the Universe which could  appear
as a result of nonperturbative effects in quantum cosmology~\cite{OPEN} may be
extremely sensitive to the properties of the theory at   nearly Planckian
densities and to the presence or absence of stationary solutions for the
distribution $P_p$. At the very least, what we have found can be considered as
a description of  rather nontrivial  properties of   hypersurfaces of a given
synchronous time in the inflationary universe. However,  we hope that some of
our results may have deeper physical significance.

\section*{Acknowledgements}

J.G.-B. is supported by a PPARC postdoctoral fellowship. He would like
to thank the warm hospitality and financial support of the Theoretical
Astrophysics Group at Fermilab, where part of this work was developed.
The work by A.L. was supported in part  by NSF grant PHY-8612280.


\newpage


\begin{thebibliography}{999}
%
%
\bibitem{Book} A.D. Linde, {\it Particle Physics and Inflationary
Cosmology} (Harwood, Chur, Switzerland, 1990).
%
\bibitem{Open} J.R. Gott, III, Nature {\bf 295}, 304 (1982);
J.R. Gott, III, and T.S. Statler, Phys. Lett. {\bf B136}, 157 (1984);
M. Sasaki, T. Tanaka, K. Yamamoto, and J. Yokoyama,
Phys. Lett. {\bf B317}, 510 (1993);  M. Bucher,
A.S. Goldhaber, and N. Turok, ``An Open Universe From Inflation,"
Princeton University preprint PUPT-1507, hep-ph/9411206  (1994);
K. Yamamoto, T. Tanaka, and M. Sasaki, Phys. Rev. {\bf D15} (1995),
pp. 2968 and 2979.
%
\bibitem{Lab} A.D. Linde,  Nucl. Phys. {\bf B372}, 421 (1992).
%
\bibitem{OMEGA} A.D. Linde, ``Inflation with Variable $\Omega$,''
Stanford University preprint SU-ITP-95-5 (1995), hep-th/9503097,
to apper in Phys. Lett. B.
%
\bibitem{JBD} P. Jordan, Nature (London) {\bf 164} (1949) 637;
Z. Phys. {\bf 157}, 112 (1959); C.H. Brans and R.H. Dicke,
Phys. Rev. {\bf 124}, 925 (1961); R.H. Dicke,
Phys. Rev. {\bf 125}, 2163 (1962); C.H. Brans,
Phys. Rev. {\bf 125}, 2194 (1962).
%
\bibitem{Olive} B. Campbell, K. Olive and A. Linde,  Nucl.Phys.
{\bf B355}, 146 (1991).
%
\bibitem{GBQ} J. Garc\'{\i}a--Bellido and
M. Quir\'os, Nucl. Phys. {\bf B368}, 463 (1992).
%
\bibitem{Polyakov}  T. Damour and A.M. Polyakov, Nucl. Phys.
{\bf B423}, 532 (1994); T. Damour and A.M. Polyakov, ``String
Theory and Gravity'', Report gr-qc/9411069 (1994).
%
\bibitem{EI} D. La and P.J. Steinhardt, Phys. Rev. Lett.
{\bf 62}, 376 (1989).
%
\bibitem{HEI} P.J. Steinhardt and  F.S. Accetta, Phys.  Rev.
Lett. {\bf 64}, 2740 (1990); J.D. Barrow and  K. Maeda, Nucl.
Phys. {\bf B341}, 294 (1990); J. Garc\'{\i}a--Bellido and M.
Quir\'{o}s,  Phys. Lett. {\bf B243}, 45 (1990).
%
\bibitem{LiddleLyth} A.R. Liddle and D.H. Lyth,  Phys. Lett.
{\bf B291}, 391 (1992).
%
\bibitem{CrittStein} R. Crittenden and P.J. Steinhardt,  Phys.
Lett.  {\bf B293}, 32 (1992).
%
\bibitem{ExtChaot} A.D. Linde, Phys.
Lett. {\bf B238}, 160 (1990).
%
\bibitem{LLM}  A.D. Linde and A. Mezhlumian,  Phys. Lett.
{\bf B307},  25  (1993);  A.D. Linde, D.A. Linde,  and
A. Mezhlumian,  Phys. Rev. {\bf D49},  1783  (1994).
%
\bibitem{Vil} A. Vilenkin, Phys. Rev. Lett. {\bf 74}, 846
(1995).
%
\bibitem{GBL} J. Garc\'{\i}a--Bellido and A.D. Linde,
Phys. Rev. {\bf D51}, 429 (1995).
%
\bibitem{GBLL} J. Garc\'{\i}a--Bellido, A.D. Linde, and D.A.
Linde, Phys. Rev. {\bf D50}, 730 (1994).
%
\bibitem{JGB} J. Garc\'{\i}a--Bellido,
Nucl. Phys. {\bf B423}, 221 (1994).
%
\bibitem{Hybrid} A.D. Linde,  Phys. Rev. {\bf D49}, 748 (1994).
%
\bibitem{Liddle} A.M. Laycock and  A.R. Liddle, Phys. Rev.
{\bf D49}, 1827 (1994).
%
\bibitem{OPEN} A.D. Linde, D.A. Linde  and
A. Mezhlumian, Phys. Lett. {\bf B345}, 203 (1995).
%
\bibitem{TEGP} C.M. Will, {\it Theory and Experiment in
Gravitational Physics} (Cambridge U.P., 1993).
%
\bibitem{PNB} J.A. Casas, J. Garc\'{\i}a--Bellido and
M. Quir\'os, Phys. Lett. {\bf B278}, 94 (1992); F.S. Accetta,
L.M. Krauss and P. Romanelli, Phys. Lett. {\bf B248}, 146 (1990).
%
\bibitem{STAR} A.A. Starobinsky, J. Yokoyama, ``Density
Fluctuations in Brans--Dicke Inflation'', preprint
astro-ph/9502002, to appear in the proceedings of the Fourth
Workshop on General Relativity and Gravitation, eds. K. Maeda
{\it et. al.}, Kyoto, Japan (1994).
%
\bibitem{DIL} J.A. Casas, J. Garc\'{\i}a--Bellido and
M. Quir\'os, Nucl. Phys. {\bf B361}, 713 (1991);
L.J. Garay and J. Garc\'{\i}a--Bellido, Nucl. Phys. {\bf B400},
416 (1993).
%
\bibitem{KriveLinde}  I.V. Krive and A.D. Linde,
Nucl. Phys. {\bf B117}, 265 (1976).
%
\bibitem{OPEN2} A.D. Linde, D.A. Linde, and A. Mezhlumian, in preparation.
%
\bibitem{HH} A. Vilenkin, Phys. Lett. {\bf B117},  25 (1982); J.B. Hartle and
S.W. Hawking, Phys. Rev. {\bf D28}, 2960 (1983).
%
\bibitem{Creation} A.D. Linde, JETP {\bf 60},  211 (1984); Lett. Nuovo
Cim. {\bf 39},  401 (1984); Ya.B. Zeldovich and A.A. Starobinsky, Sov. Astron.
Lett. {\bf 10},  135 (1984); V.A. Rubakov, Phys. Lett. {\bf 148B},  280 (1984);
A. Vilenkin, Phys. Rev. {\bf D30},  549 (1984).
%
\bibitem{Tunn} A. Starobinsky, in H.J. de Vega and N. Sanchez, Eds.,
{\it Current Topics in Field Theory,
Quantum Gravity and Strings,} Lecture Notes in Physics {\bf 206,}
(Heidelberg: Springer)(1986);
A. Goncharov and A. Linde,   Sov. J. Part. Nucl.
{\bf 17,} 369 (1986);
A. Linde,   Nucl. Phys. {\bf B372,} 421 (1992).
%
\bibitem{HR} S.W. Hawking  and S.F. Ross, ``Duality between Electric
and Magnetic Black Holes'', preprint DAMTP/R-95/8, hep-th/9504019 (1995).
%
\bibitem{JGBDW} J. Garc\'\i a--Bellido and D. Wands,
``General Relativity as an Attractor of Scalar--Tensor Stochastic
Inflation'', preprint SUSSEX-AST-95/3-1, gr-qc/9503049 (1995).
%


\end{thebibliography}
\end{document}